\documentclass[12pt]{article}
\usepackage{amssymb,amsmath}
\usepackage{natbib}
\usepackage{graphicx}
\usepackage{epsfig}

\begin{document}

\title{Spiral patterns in planetesimal circumbinary disks}

\author{Tatiana V. Demidova, Ivan I. Shevchenko\/\thanks{E-mail:~iis@gao.spb.ru} \\
Pulkovo Observatory of the Russian Academy of Sciences \\
Pulkovskoje ave. 65, St.\,Petersburg 196140, Russia}

\date{}

\maketitle

\begin{center}
Abstract
\end{center}

\noindent Planet formation scenarios and the observed planetary
dynamics in binaries pose a number of theoretical challenges,
especially in what concerns circumbinary planetary systems. We
explore the dynamical stirring of a planetesimal circumbinary disk
in the epoch when the gas component disappears. For this purpose,
following theoretical approaches by Heppenheimer (1978) and
Moriwaki \& Nakagawa (2004), we develop a secular theory for the
dynamics of planetesimals in circumbinary disks. If the binary is
eccentric and its components have unequal masses, a spiral density
wave is generated, engulfing the disk on the secular timescale,
which may exceed $10^7$~yr, depending on the problem parameters.
The spiral pattern is transient; thus, its observed presence may
betray system's young age. We explore the pattern both
analytically and in numerical experiments. The derived analytical
spiral is a modified {\it lituus}; it matches the numerical
density wave in the gas-free case perfectly. Using the SPH scheme,
we explore the effect of residual gas on the wave propagation.

\bigskip

\noindent Key words: binary stars -- exoplanets -- planetesimals
-- astrophysical disks -- circumbinary planetary systems.

\newpage

\section{Introduction}

Planet formation scenarios and the observed planetary dynamics in
binaries pose a number of theoretical challenges, especially in
what concerns circumbinary planetary systems. The orbital motion
of the central binary induces the formation of global structures
in the gas-dust protoplanetary disk. In particular, a cavity is
formed in the central part of the circumbinary disk; though, the
cavity can be penetrated by streams of matter, maintaining the
accretion activity of the binary components. Numerical simulations
of the gas accretion in such systems were performed in
\citep{AL96,BB97,SG07} using the SPH method, and in
\citep{GK02,O05,H10,K10} using finite-difference schemes.

In a general theoretical context, various spiral wave modes, which
may emerge due to tidal interaction with planets in gaseous
circumstellar disks, were revealed in \cite{LO98}. An empirical
five-parameter fitting relation describing the spiral pattern
excited by a planet in a gaseous disk was proposed in
\cite{MGH12}. In numerical simulations of gaseous protoplanetary
disks in binary systems, as well as in simulations of disks with
embedded planets, a variety of spiral patterns were found out to
form \citep{N00,KN08,P08,M09,MB12,PM13}. Recent ALMA observations
provide remarkable examples of spiral structures in gaseous disks
excited by embedded planets \citep{C14} or by central stellar
binaries \citep{T14}.

Spiral density waves may as well emerge in the inner zones of cool
accretion disks around components of semidetached binary systems.
The possibility of observations of such patterns, obtained in
three-dimensional gas-dynamical simulations, was discussed in
\cite{BB04}. The pattern formation in such disks is due to the
differential orbital precession of the accreting particles, the
precession being caused by the perturbations from the donor star.

Concerning the gas-free planetesimal disks, it is well known that
a spiral structure can form, if a planet is present in such a
disk~\citep{W05,MW09,MKW14}. \cite{W05} showed that a tightly
wound spiral starts to propagate through the disk, if a planet is
introduced in it. As in the theory of \cite{BB04}, the spiral is
generated by the differential precession of the neighboring
particle orbits. To estimate the perturbation effect of a planet
on the orbits of the planetesimals, \cite{W05} used a theory of
the secular variation of the complex eccentricity coupled with the
variation of the longitude of pericenter, using the Lagrange
planetary equations. He applied this secular theory to explain a
spiral pattern resolved by \cite{C03} in the image of the
HD~141569 disk, and to interpret the results of a numerical
simulation of the planetesimal disk evolution in presence of a
giant planet. An important effect, considered by \cite{MW09}, is
the dynamical heating of the disk as a result of the spiral wave
propagation, caused by a planet introduced in the disk: the
planetesimal velocity dispersion increases. This phenomenon,
hindering the planetesimal coalescence, may affect the further
process of planetary formation. Of course, for the circumbinary
case, considered below, a similar effect may be of even greater
importance.

The protoplanetary disks strongly radiate in the infrared and
submillimeter spectral ranges, because they contain small-sized
dust particles. The fine dust fraction decreases with time due to
the particles coagulation; on the other hand, numerical
simulations show that the collisional destruction of planetesimals
is also an effective process, causing the dust to persist over
millions of years \citep{B09}. Many images of gas-rich,
transitional and debris disks have been obtained; see, e.g.,
\citep{P99,G05,K05,MG12,MS12,K12}. In these images, global
structures were detected, namely: spiral arms
\citep{G01,H11,C14,T14}, ring-like gaps \citep{W99}, bright rings
\citep{K05}, inner holes \citep{MG12,MS12}, warps \citep{H00},
density clamps \citep{G98}.

The formation of spiral waves in the gaseous disk depends on the
viscous gas properties. When the gaseous component disappears, the
accretion stops, and any further structures are formed in the
planetesimal ensemble. In this paper, we explore the dynamical
stirring of a planetesimal circumbinary disk in the epoch when the
gas component disappears. We show that the orbital precession of
the particles forming the disk generates prominent spiral
patterns, and we describe these patterns analytically.

In our analysis, we assume that the timescale for the secular wave
propagation is greater than the lifetime of the gas-rich disk (and
by far greater than the characteristic timescales of gas depletion
and planetesimal formation). Whether this assumption is realistic
for any particular system, depends on the values of a number of
parameters, which we identify. We demonstrate that the secular
wave propagation timescale can be, in principle, long enough for
our assumption to be valid.

It should be emphasized that, in this paper, we consider mostly
the {\it circumbinary} disks (in which the planetesimals are
moving around both stars situated at the disk center; thus, the
disk is an outer one with respect to the binary). {\it
Circumstellar} disks, in which the planetesimals are moving around
one of the binary components (thus, the disk is an inner one), are
discussed only in brief (in Section~\ref{sec_cdc}).

The paper is structured as follows. In Section~\ref{sec_sth}, we
consider a secular theory for the dynamics of planetesimals in
circumbinary disks, and derive necessary formulas for the secular
evolution of the planetesimal eccentricity and longitude of
pericenter. An analytical expression describing the geometry of
the spiral density wave is given. In Section~\ref{sec_ts}, we
write down a formula for the secular timescale of the wave
propagation, and discuss whether this timescale can exceed the
lifetime of a gas-rich disk in realistic systems. In
Section~\ref{sec_dft}, the radial (with respect to the central
binary) secular ``oscillations'' of the planetesimal eccentricity
are investigated analytically and compared to known results of
numerical simulations; besides, we consider the time evolution of
individual orbits and, again, make comparisons with known results
of numerical simulations. In Section~\ref{sec_cdc}, a secular
theory for the circumstellar case is briefly discussed. In
Section~\ref{sec_cpd}, we explore correspondence between the
analytical spiral pattern and the numerical-experimental density
waves in the circumbinary case. Conclusions are summarized in
Section~\ref{sec_concl}.

\section{The secular theory in the circumbinary case}
\label{sec_sth}

In this section, we develop a secular theory for the dynamics of
planetesimals in circumbinary disks,  following theoretical
approaches by \cite{H78} and \cite{MN04}. A secular perturbation
theory, providing analytical formulas for the particle's
eccentricity $e$ and the longitude of periastron $\varpi$, was
derived by \cite{H78} (see also \citealt{WMC98,T06}) for the
circumstellar (circumprimary or circumsecondary) case. It was used
to analytically describe, how a circumstellar disk of a young star
is stirred by a companion star. On the other hand, the
circumbinary case was considered by \cite{MN04}. They presented
approximate differential equations for the secular evolution of
the eccentricity vector; see Appendix in \cite{MN04}. We combine
the approaches of \cite{H78} and \cite{MN04} to derive necessary
explicit analytical formulas for the secular evolution of the
planetesimal eccentricity and longitude of pericenter.

Hereafter we adopt the barycentric frame; $m_1$ and $m_2$ are the
masses of the binary components (we set $m_1 \ge m_2$),
$a_\mathrm{b}$ is the binary semimajor axis, $e_\mathrm{b}$ is the
binary eccentricity, $a$ is the semimajor axis of the planetesimal
orbit. The masses are measured in the solar units, distances in
astronomical units (AU), time in years. Thus, the gravitational
constant $G$ is equal to $4\pi^2$.

The {\it hierarchical} three-body problem ``binary--planetesimal''
is considered: the distance of the planetesimal from the binary
barycenter is assumed to be much greater than the size of the
binary. In fact, it is superfluous to consider a non-hierarchical
case, because a large central chaotic circumbinary zone exists at
all eccentricities of the planetesimal, if $\mu \gtrsim 0.05$
\citep{S15}, where $\mu = m_2 / (m_1 + m_2)$ is the mass parameter
of the binary. This relative mass threshold has an important
physical meaning \citep{S15}: above it, the tertiary
(planetesimal), even starting from small eccentricities, can
diffuse, following the sequence of the overlapping $p$:1 mean
motion resonances between the tertiary and the binary, up to
ejection from the system; close encounters with other bodies are
not required for the escape. (Note that, on the other hand, in the
case of inner circumstellar orbits, stable orbits always exist
inside the Hill spheres of the binary components.) In what
follows, we assume that $\mu \gtrsim 0.05$.

We use an averaged perturbing function expansion presented in
\cite{MN04} for the circumbinary case. This is a power-law
expansion in the ratio of the binary and planetesimal semimajor
axes and in the eccentricities. The expansion is up to the third
order in the ratio of the semimajor axes and up to the fourth
order in the eccentricities, inclusive. Note that the ratio of the
binary and planetesimal semimajor axes is assumed to be small,
since the hierarchical problem is considered. We integrate
analytically the corresponding equations of motion (see
Equations~(A7) in \citealt{MN04}), and thus straightforwardly
deduce formulas for the secular evolution of the eccentricity $e$
and the longitude of periastron $\varpi$ of the circumbinary
planetesimal. They turn out to be very similar to those in the
circumstellar case, presented in \cite{H78}, and read:

\begin{equation}
e = e_\mathrm{max} \left| \, \sin \frac{ut}{2} \right| ,
\label{eccen}
\end{equation}

\begin{equation}
\tan \varpi = -\frac{\sin ut}{1-\cos ut} ,
\label{periastron}
\end{equation}

\noindent where $t$ is time,

\begin{equation}
u = \frac{3 \pi}{2} \frac{m_1 m_2}{(m_1 + m_2)^{3/2}}
\frac{a_\mathrm{b}^2}{a^{7/2}} \left(1 + \frac{3}{2}
e_\mathrm{b}^2 \right) ,
\label{prec}
\end{equation}

\noindent and $e_\mathrm{max}=2e_\mathrm{f}$, where $e_\mathrm{f}$
is the so-called forced eccentricity:

\begin{equation}
e_\mathrm{f} = \frac{5}{4} \frac{(m_1 - m_2)}{(m_1 + m_2)}
\frac{a_\mathrm{b}}{a} e_\mathrm{b} \frac{\left( 1 + \frac{3}{4}
e_\mathrm{b}^2 \right)} {\left( 1 + \frac{3}{2} e_\mathrm{b}^2
\right)} .
\label{eforced}
\end{equation}

\noindent Using a new variable, namely, $y = \frac{u t}{2}$, we
rewrite Equation~(\ref{periastron}) in the form

\begin{eqnarray}
&& \mbox{if} \ y \geq -\pi \ \mbox{and} \ y \leq -\frac{\pi}{2} , \ \mbox{then} \ \varpi = y + 5 \frac{\pi}{2} ; \nonumber \\
&& \mbox{if} \ y \geq -\frac{\pi}{2} \ \mbox{and} \ y \leq 0 , \ \mbox{then} \ \varpi = y + \frac{\pi}{2} ; \nonumber \\
&& \mbox{if} \ y \geq 0 \ \mbox{and} \ y \leq \frac{\pi}{2} , \ \mbox{then} \ \varpi = y + 3 \frac{\pi}{2} ; \nonumber \\
&& \mbox{if} \ y \geq \frac{\pi}{2} \ \mbox{and} \ y \leq \pi , \ \mbox{then} \ \varpi = y - \frac{\pi}{2} .
\label{varpi}
\end{eqnarray}

\noindent Thus, $u$ can be regarded as a ``precession rate''
(though in a modified fashion) of an individual orbit.

Based on the analytical dependences of the eccentricity and the
longitude of periastron on the semimajor axis
(Equations~(\ref{eccen}) and (\ref{varpi})), let us proceed to
describe the circumbinary disk structure, emerging due to the
secular perturbations from the central binary. Namely, let us
derive an analytical formula for the spiral pattern in the
gas-free case. This can be done by tracing the radial distance $r$
of the apocenter at the maximum (in the course of its secular
evolution) eccentricity of particle's orbit, as a function of the
polar angle $\theta$. The apocenters are attained at $\varpi =
\pi$ mod $2\pi$, i.e., at the polar angle $\theta = \pi$ mod
$2\pi$. At these points, the analytical solution is known
(Equations~(\ref{eccen}) and (\ref{varpi})). Taking into account
that at $\theta = 0$ mod $2\pi$ the spiral corresponds to
particle's minimum (zero) eccentricity, and interpolating between
the consecutive points $\theta = \pi n$ ($n = 1, 2, 3, \dots$),
one arrives at

\begin{equation}
r(\theta, t) = \left( \frac{A t}{\theta} \right)^{2/7} + B (1 -
\cos \theta) , \label{sp_formula}
\end{equation}

\noindent where the constants

\begin{equation}
A =\frac{3\pi}{2} \frac{m_\mathrm{1}
m_\mathrm{2}}{(m_\mathrm{1}+m_\mathrm{2})^{3/2}} a_\mathrm{b}^2
\left( 1+\frac{3}{2}e_\mathrm{b}^2 \right), \quad B = \frac{5}{4}
\frac{m_\mathrm{1}-m_\mathrm{2}}{m_\mathrm{1}+m_\mathrm{2}}
a_\mathrm{b} e_\mathrm{b} \frac{\left( 1+\frac{3}{4}e_\mathrm{b}^2
\right)} {\left( 1+\frac{3}{2}e_\mathrm{b}^2 \right)}.
\label{sp_formula_AB}
\end{equation}

\noindent Thus, the pattern represents a shifted ``power-law
spiral''. The main term, proportional to $\theta^{-2/7}$,
represents a generalized {\it lituus} (a crook). (Note that the
classical lituus is a little bit different, being given by $r
\propto \theta^{-1/2}$.) The shifting term $\propto (1 - \cos
\theta)$ represents the {\it cardioid}.

\section{The secular timescale}
\label{sec_ts}

The observability of the derived secular spiral pattern depends on
an interplay of several dynamical and physical timescales. As
determined from observations of stars in young clusters and
associations, the protoplanetary disks start their evolution in a
gas-rich phase, which lasts some $\sim 10^6$--$10^7$~yr
\citep{HLL01,WC11}. Then, in a relatively short-lived ($\sim
10^5$~yr) ``transitional disk'' phase \citep{CPS07,WC11}, gas is
lost; the pure debris disk is left, consisting of planetesimals,
possibly with planets or planetary embryos already embedded. The
planetesimals form (by dust accretion) already in the first
gas-rich phase, on a short timescale $\sim 10^4$~yr \citep{W97};
the planets or planetary embryos may also form in the first
(gas-rich) phase (see, e.g., \citealt{KI96,INE00}).

In our analysis, we assume that the timescale for the secular wave
propagation across the disk is greater than the lifetime of the
gas-rich disk, and by far greater than the characteristic
timescales of gas depletion and planetesimal formation.

From Equations~(\ref{sp_formula}) and ~(\ref{sp_formula_AB}) one
can readily find the {\it secular timescale} $T_\mathrm{s}$,
namely, the time needed for the density wave to propagate across
the disk. For this purpose, we set $\theta = \pi$ (the value
opposite to the singular value $\theta = 0$), and write down the
time $t=T_\mathrm{s}$ as a function of $r=r_\mathrm{disk}$ (the
disk's outer edge radius):

\begin{equation}
T_\mathrm{s} = \frac{\pi}{A} (r_\mathrm{disk} - 2 B)^{7/2} \approx
\frac{\pi}{A} r_\mathrm{disk}^{7/2} . \label{Ts}
\end{equation}

\noindent The approximate equality is valid because the problem is
hierarchical.

Thus, the secular timescale depends on a number of parameters,
and, first of all, rather strongly on the disk radius. It
increases sharply with $r_\mathrm{disk}$ and decreases (rather
moderately) with $a_\mathrm{b}$, $\mu$, and $m_1$. The binary
eccentricity $e_\mathrm{b}$ is relatively unimportant.

For Kepler-16 ($m_1=0.69$~$M_\odot$, $m_2=0.2029$~$M_\odot$,
$a_\mathrm{b}=0.2243$~AU, $e_\mathrm{b}=0.159$), if one sets
$r_\mathrm{disk}=30$~AU, one has $T_\mathrm{s} = 1.1 \cdot
10^7$~yr. For an ``M-dwarf--brown dwarf'' system
($m_1=0.5$~$M_\odot$, $m_2=0.025$~$M_\odot$, $a_\mathrm{b}=1$~AU,
$e_\mathrm{b}=0.1$), if one sets $r_\mathrm{disk}=100$~AU, one has
$T_\mathrm{s} = 2.0 \cdot 10^8$~yr. Therefore, the secular
timescale can vary significantly from system to system.

For the secular spiral be observable, one must assume that the
timescale for the secular wave propagation is greater than the
lifetime of the gas-rich disk ($\sim 10^6$--$10^7$~yr, see
Introduction). Whether this assumption is realistic for any
particular system, depends on the values of the problem
parameters: $r_\mathrm{disk}$, $a_\mathrm{b}$, $\mu$, and $m_1$.
As we have just seen, the secular wave propagation timescale can
be, in principle, long enough: it may well exceed $10^7$~yr,
depending on the problem parameters.

\section{Radial ``oscillations'' of eccentricity and time evolution of
individual orbits}
\label{sec_dft}

Numerical simulations of the dynamical stirring of planetesimal
disks on secular timescales, in various problem settings, were
performed in \citep{MN04,M12a,M12b,P12}. In particular, graphs of
the eccentricity and longitude of periastron of a circumbinary
particle as a function of its semimajor axis were constructed
numerically. Let us show that such graphs can be as well
constructed analytically using the theory described above.

\begin{figure}[ht!]
\begin{center}
\includegraphics[width=14cm]{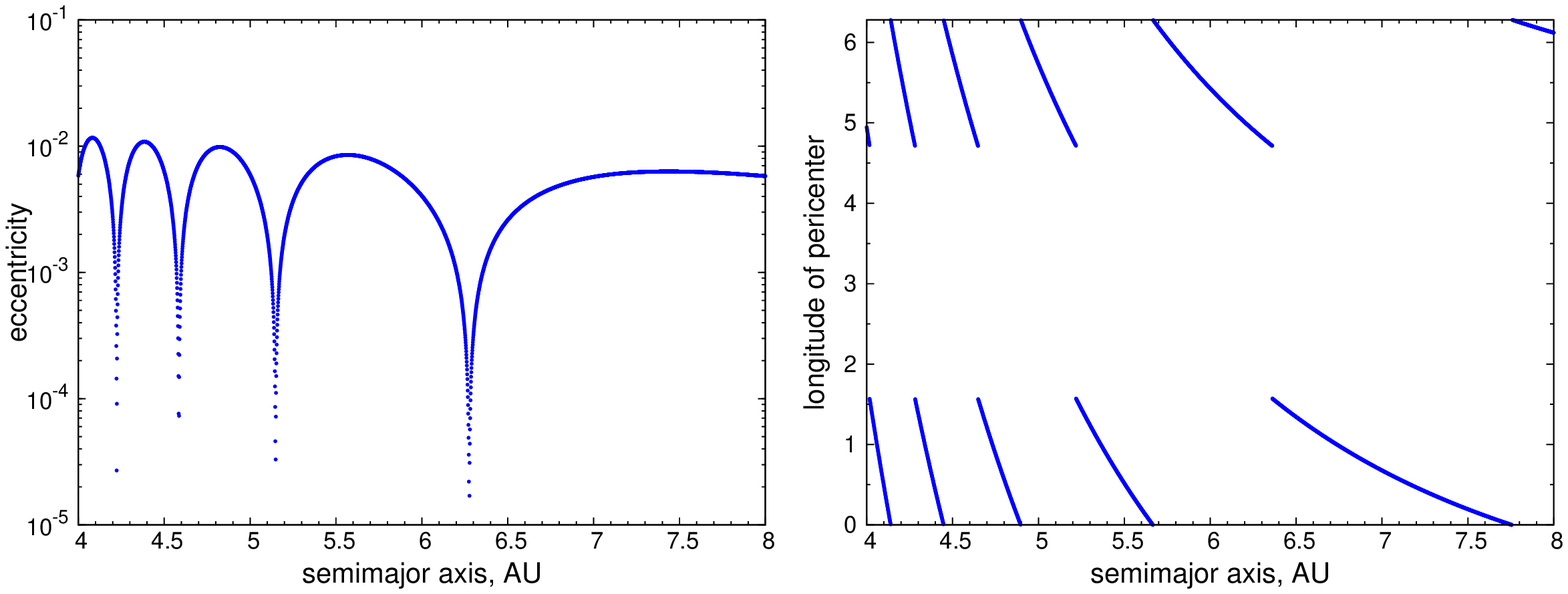}
\end{center}
\caption{\small The secular eccentricity ``waves'' (left) and the
secular behavior of the longitude of pericenter (right) of
planetesimals in a model disk, as a function of the semimajor axis
at a fixed time. For the model parameters see the text.}
\label{fig1}
\end{figure}

As an actual binary example, we consider {\it Kepler}-16. This
binary is remarkable to host the first ever discovered
circumbinary planet in a double main-sequence star system. The
binary's parameters are: $m_1 = 0.69 M_\odot$, $m_2 = 0.2026
M_\odot$, $e_\mathrm{b} = 0.159$, $a_\mathrm{b} = 0.2243$~AU
\citep{D11}. Analytically constructed diagrams, as given by
Equations~(\ref{eccen}) and (\ref{varpi}) at $t=10^5$~yr are shown
in Figure~\ref{fig1}. Radial ``waves'' of eccentricity, well-known
from numerical diagrams in \citep{MN04,M12a,M12b,P12}, are clearly
recognizable.

Our diagrams match exactly those constructed in \citep{MN04,M12b}
in the models with the same parameters, when the gas component is
absent. A direct comparison of our diagrams with those constructed
in \citep{M12a,P12} in the models with the gas component shows
that the presence of gas increases the timescale of eccentricity
pumping.

An analogous approach can be used to describe secular oscillations
of a circumbinary orbit of a particle (a planetesimal or a planet)
in dependence on time, at a fixed semimajor axis. In this respect,
one can compare the theory described above with the results of
numerical simulations performed in \citep{LL13,DA13,M14} for the
planet {\it Kepler}-16b. We set the binary parameters to be the
same as cited above in this section, and $a=0.7016$~AU is fixed
for the planet. The initial conditions for the planet motion are
taken from \cite{D11}.

\begin{figure}[ht!]
\begin{center}
\includegraphics[width=14cm]{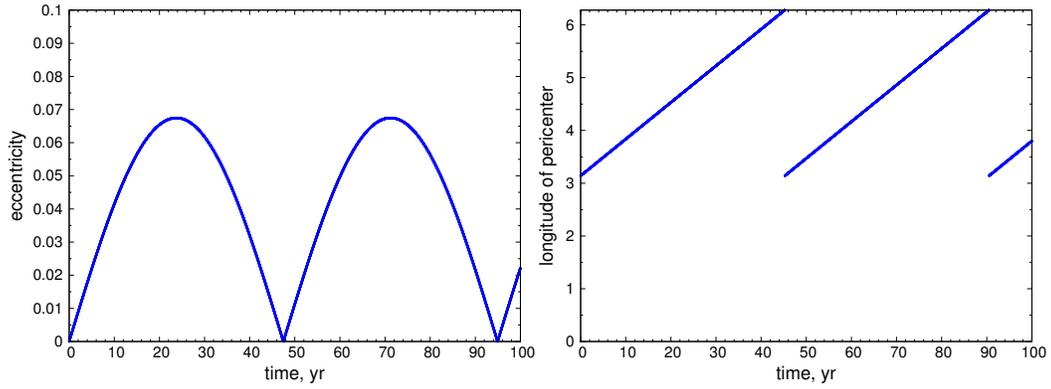}
\end{center}
\caption{\small The eccentricity secular oscillations (left) and
the longitude-of-pericenter secular rotation (right) in case of
{\it Kepler}-16b, as a function of time at a fixed semimajor
axis.}
\label{fig2}
\end{figure}

In Figure~\ref{fig2}, time dependences for the eccentricity and
the longitude of periastron, given by Equations~(\ref{eccen}) and
(\ref{varpi}), are constructed. As follows from a direct
comparison with the corresponding numerical-experimental graphs in
\citep{LL13,M14}, our secular theory formulas match these graphs
perfectly.

\section{Circumstellar disks}
\label{sec_cdc}

As opposed to the circumbinary case, the {\it circumstellar} disks
are defined as those in which the planetesimals are moving around
one of the binary components. Thus, the disk is an inner one, with
respect to the binary itself. In this section, we briefly consider
the circumstellar case.

Planetesimals orbiting a component of a binary star are subject to
perturbations due to the stellar companion. The perturbations pump
the eccentricities of planetesimals and inhibit their accumulation
process \citep{MS00,T04}. In \cite{T06}, the timescales for the
inward propagation of the orbital crossing ``wave'' are calculated
in gas-free models. In the more realistic models, the density wave
inward propagation is damped by the gas drag and collisions.

In circumstellar disks, strong spiral arms can be present, subject
to retrograde precession \citep{KN08,P08}. Oscillations of
planetesimal orbital parameters (the eccentricity and the
longitude of pericenter) as a function of time, at a fixed
semimajor axis, are presented graphically in
\citep{KN08,M09,BL10,MB12,MK12,L13} for various values of disk
parameters (mass, viscosity, radiative transfer properties).
Possible mechanisms accounting for the spiral structure in a
circumstellar gaseous disk perturbed by an outer body (planet,
star, brown dwarf) were studied in \cite{Q05}, using
two-dimensional hydrodynamic simulations. Numerically simulated
diagrams of the eccentricity and the longitude of periastron of
planetesimals as a function of the semimajor axis, at a fixed
time, are presented in \citep{MS00,P08,PL10,L13}. The choice of
parameters for the simulations usually correspond to the binary
systems $\gamma$~Cep and $\alpha$~Cen, possessing circumstellar
planets.

What are the basic expressions describing the secular evolution in
the circumstellar case? The expressions for the eccentricity and
longitude of periastron are virtually the same as in the
circumbinary case: they are given by Equations~(\ref{eccen}),
(\ref{periastron}), and (\ref{varpi}), but the formulas for the
$u$ and $e_\mathrm{f}$ parameters, entering these expressions, are
different. We quote them as given in \citep{H78,WMC98,T06}, but
arrange in our notations:

\begin{equation}
u = \frac{3 \pi}{2} \frac{m_2}{m_1^{1/2}}
\frac{a^{3/2}}{a_\mathrm{b}^3} \left(1 - e_\mathrm{b}^2
\right)^{-3/2} , \label{prec_inner}
\end{equation}

\begin{equation}
e_\mathrm{f} = \frac{5}{4} \frac{a}{a_\mathrm{b}}
\frac{e_\mathrm{b}} {\left( 1 - e_\mathrm{b}^2 \right)} .
\label{eforced_inner}
\end{equation}

\noindent Here $m_1$ is the primary mass (around which the
particle orbits), and $m_2$ the perturbing mass; $m_1 > m_2$. Note
that in the treatment by \cite{H78} the perturbing function is not
expanded in the eccentricities, but solely in the ratio of the
planetesimal and binary semimajor axes (up to the third order
inclusive). The ratio of the planetesimal and binary semimajor
axes is assumed to be small, since the hierarchical problem is
considered.

Thus, the secular evolution in the circumstellar case is described
analytically by Equations~(\ref{eccen}), (\ref{periastron}),
(\ref{varpi}), (\ref{prec_inner}), and (\ref{eforced_inner}).

\section{Patterns in circumbinary disks}
\label{sec_cpd}

Let us consider in more detail the formation of patterns in
circumbinary disks. As it is clear from Figure~\ref{fig1}, the
formation of secular structures in the disk is inevitable, because
the maxima of eccentricities correspond to the particles groupings
at apocenters, due to the low velocities of particles there.
Therefore, the disk radial densities are non-uniform at any fixed
time.

\begin{figure}
\begin{center}
\includegraphics[width=12cm]{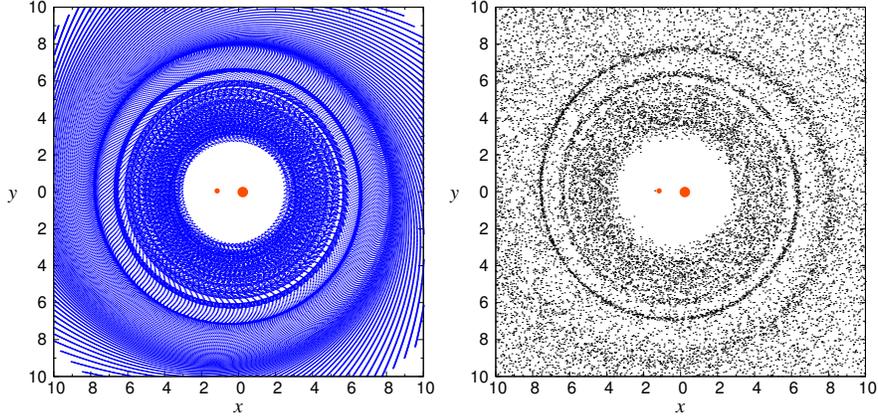}
\end{center}
\caption{\small A spiral pattern in a model planetesimal disk. The
set of precessing orbits, constructed analytically (left), and the
corresponding SPH simulation (right). The model parameters:
$m_1=M_\odot$, $m_2 = 0.2 M_\odot$, $e_\mathrm{b}=0.4$, $a=1$~AU,
$t=10^4$~yr.}
\label{fig3}
\end{figure}

Let us provide an illustrative example. Fixing the time of
evolution to $t=10^4$~yr, we calculate analytically the evolved
planetesimal orbits in the semimajor axis $a$ interval from $3
a_\mathrm{b}$ to $16 a_\mathrm{b}$ with the distance step $\Delta
a = 0.01 a_\mathrm{b}$. (The chosen inner radius corresponds
roughly to the typical size of the circumbinary chaotic zone, see
\citealt{S15}.) The planetesimal orbits at each $a$ are shown by
dots, with the time step equal to $0.01$ of the planetesimal
orbital period. The binary parameters are set to be $m_1=M_\odot$,
$m_2 = 0.2 M_\odot$, $e_\mathrm{b}=0.4$, $a_\mathrm{b}=1$~AU; this
choice roughly corresponds to the ``most eccentric'' model in
\cite{MN04}. The resulting diagram is presented in
Figure~\ref{fig3}a.

Besides, we have performed an SPH-code numerical simulation,
corresponding to this model (i.e., the parameters and initial
conditions are the same). The SPH-code realized in \citep{S96} has
been used. The number of particles is set to $30000$. The kernel
for smoothing the hydrodynamic quantities is chosen in the form of
a spline \citep{MG83}, and the smoothing length is set to be
constant. On starting the simulation, the SPH particles are placed
on circular orbits; the velocities are Keplerian. The particles
have equal sizes and masses, and are distributed in accordance
with the surface density profile $\Sigma \sim r^{-1}$. The total
mass of the planetesimal disk is assumed to be negligible compared
to that of the binary; therefore, the self-gravity of the disk is
neglected. In the gas-free case, the system of equations in the
applied SPH-code reduces to the equations of motion of the
particles in the gravitational potential of the central binary.
The results of the simulations in the gas-free case are presented
in Figure~\ref{fig3}b.

In the both panels of Figure~\ref{fig3}, spiral patterns are
evident. They are tightly bound in the direction of the companion
motion. Note a close resemblance of the patterns in the panels,
though panel~{\it a} is constructed using an analytical theory,
whereas panel~{\it b} is a pure simulation.

\begin{figure}
\begin{center}
\includegraphics[width=8cm]{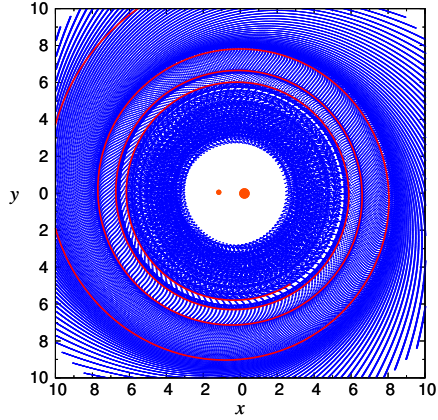}
\end{center}
\caption{\small The plot of Figure~\protect\ref{fig3}a with an
analytical spiral curve superimposed. The latter is given by
Equation~(\protect\ref{sp_formula}) and is depicted by a thick
solid line.}
\label{fig4}
\end{figure}

In a separate figure (Figure~\ref{fig4}), we illustrate
graphically how well our analytical spiral, given by
Equation~(\ref{sp_formula}), matches the density wave emerging in
the numerical simulations in the gas-free case. Namely, we
superimpose the analytical spiral on the numerical-experimental
plot taken from Figure~\ref{fig3}a. In Figure~\ref{fig4}, the
analytical curve is depicted by a thick solid line. A good
agreement between the analytical spiral pattern and the numerical
density wave is evident.

Consider now a planetesimal disk containing gas. Gas drag acting
on a planetesimal depends on gas and planetesimal parameters
\citep{W73,W77}. We consider large particles and assume that the
free path of gas molecules is less than the planetesimal radius.
The planetesimals are subject to gas drag; to calculate the effect
of gas drag, we complement the system of equations in the SPH-code
with the continuity equation for gas and the ideal gas equation of
state in the isothermal case; namely, gas pressure $P = c^2
\rho_\mathrm{g}$, where $c$ is the sound speed, $\rho_\mathrm{g}$
is gas density. To simulate gas presence in the planetesimal disk,
we add an extra acceleration term

\begin{equation}
\frac{d\vec{v}_\mathrm{p}}{dt} = - \frac{3}{8} C_\mathrm{D}
\frac{\rho_\mathrm{g}}{\rho_\mathrm{p} s} \vert v_\mathrm{p} -
v_\mathrm{g} \vert (\vec{v_\mathrm{p}}-\vec{v_\mathrm{g}})
\end{equation}

\noindent in the equations of motion of planetesimals, like it is
done in \citep{W73,W77}. Here $\rho_\mathrm{g}$ and
$\rho_\mathrm{p}$ are gas and planetesimal densities, $s$ is the
radius of a planetesimal, $v_\mathrm{p}$ and $v_\mathrm{g}$ are
planetesimal and gas velocities. The dimensionless coefficient
$C_\mathrm{D} = 0.44$, as in \citep{W73} for a regime with the
Reynolds number $\mathrm{Re} > 800$.

In our model, the sound speed is $1.2$~km/s, corresponding to the
temperature $\approx 100$~K; this gives the disk's effective
semithickness $h = H/r = 0.06$ at $r=1$~AU, where $H$ is the
disk's semithickness, and $r$ is the distance from the center of
mass. (The disk's semithickness is controlled by gas temperature,
namely, $H = 2^{1/2} c / \Omega$, where $\Omega$ is the Keplerian
angular velocity.) We describe the gaseous disk's effective
viscosity in the $\alpha$-parametrization \citep{SS73,P81}:
$\nu=\alpha c H$, where $\alpha$ is the Shakura--Sunyaev viscosity
parameter; $\alpha \approx 0.03$, like in \cite{AL96}. The model's
other parameters are as follows: the planetesimal radius $s=1$~km,
the planetesimal density $\rho_\mathrm{p} = 1$~g~cm$^{-3}$, gas
total mass in the disk $M_\mathrm{g}=10^{-2}$~$M_{\odot}$.

Our SPH simulations of the gas-containing disk show that the
presence of residual gas slows down the wave propagation, the wave
pattern remaining, however, self-similar. If the presence of gas
is great enough, the spiral pattern does not emerge altogether.
Note that, since the work of \cite{MS00}, it is well known that
the local gas damping, if high enough, causes an alignment of the
apsidal lines of circumbinary orbits of equal-sized planetesimals.
Therefore, increasing the gas amount must suppress the secular
spiral structure formation in the planetesimal disk.

\begin{figure}
\begin{center}
\includegraphics[width=12cm]{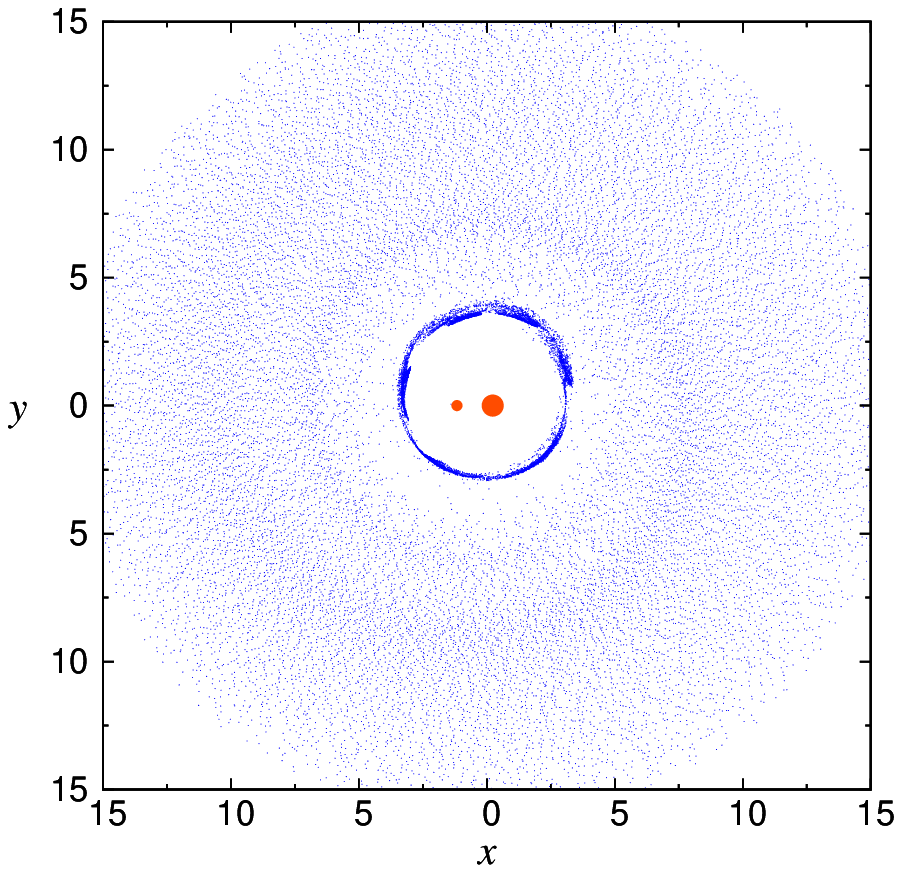}
\end{center}
\caption{\small The same as Figure~\protect\ref{fig3}b, but the
disk has a gaseous component, as described in the text.}
\label{fig5}
\end{figure}

We limit our analysis of gas-containing disks by these short
considerations. The problem of pattern formation in the
two-component (gas plus planetesimals) media deserves a detailed
separate study, and we leave it for the future work. Let us give
just one illustrative example (Figure~\ref{fig5}). The initial
data for the planetesimal disk are the same as used for
construction of Figure~\ref{fig3}b, but a small amount of gas is
introduced (the model parameters are as described above). As
follows from the simulations, it requires $10^4$~yr to form the
same spiral pattern as emerging in the gas-free case in $8 \cdot
10^3$~yr. In Figure~\ref{fig5}, it is evident that gas amount is
insufficient in the given model to suppress the spiral pattern
completely. On the other hand, one can see that an elliptical
structure is also formed (at the inner edge of the disk), with the
major axis perpendicular to the apsidal line of the central
binary; this is in accord with the theory by \cite{MS00} on the
alignment of planetesimal orbits due to gas drag.

Note that the disk self-gravity is ignored in our simulations. The
self-gravity effect depends on the masses of planetesimals and on
their spatial concentration; thus, we assume that the masses
and/or concentration are small enough. However, one may speculate
that the self-gravity effect, when non-negligible, is generally to
suppress the spiral pattern, because, as follows from simulations
\citep{M09}, it leads to low eccentricities of the disk particles.

Concerning the circumstellar case, the ``secular'' spiral pattern
is basically analogous to that derived above in this section for
the circumbinary case. A detailed analysis for the circumstellar
case, alongside with comparisons with the numerous simulation
results cited in Section~\ref{sec_cdc}, will be performed
elsewhere.

\section{Conclusions}
\label{sec_concl}

Thus, we have combined the approaches of \cite{H78} and
\cite{MN04} to derive explicit analytical formulas for the secular
evolution of the eccentricities and longitudes of pericenters of
planetesimals in circumbinary disks, when the gas component
disappears. We have shown that, if the binary is eccentric and its
components have unequal masses, spiral density waves are
generated. They engulf the disk on the secular timescale, which
may well exceed $10^7$~yr, depending on the problem parameters.
Being transient (on the secular timescale), their observed
presence may betray system's young age.

Our analytical results, given by
Equations~(\ref{eccen})--(\ref{varpi}) and
Equations~(\ref{prec_inner})--(\ref{eforced_inner}), explain
qualitatively and quantitatively many numerical-expe\-ri\-men\-tal
diagrams, such as presented in
\citep{MS00,MN04,T06,P08,PL10,M12a,M12b,P12,L13} for the
eccentricity and the longitude of periastron as a function of the
semimajor axis at a fixed time, as well as presented in
\citep{M09,BL10,MK12,LL13,L13,M14} for the eccentricity and the
longitude of periastron as a function of time at a fixed semimajor
axis.

For the gas-free case, we have derived an analytical formula for
the spiral pattern; this formula describes a modified ``lituus''
(a shifted power-law spiral). Its form is in perfect agreement
with our results of modeling the precessing planetesimal orbits,
as well as our results of direct SPH simulations performed for the
same disk parameters and initial conditions. Using the SPH scheme,
we have observed that the effect of residual gas is to slow down
the spiral wave propagation.

In our analysis, we have assumed that the timescale for the
secular wave propagation is greater than the lifetime of the
gas-rich disk ($\sim 10^6$--$10^7$~yr). Whether this assumption is
realistic for any particular system, depends on the values of a
number of parameters, which we have identified. We have
demonstrated that the secular wave propagation timescale can be,
in principle, long enough: it may exceed $10^7$~yr, depending on
the problem parameters.

One should underline that the circumbinary spiral pattern emerges
if the binary is eccentric and its components have unequal masses.
If any of these two conditions is violated, concentric circular
waves (``secular circles'') are formed instead of the spiral wave,
as it is evident from the results of numerical simulations
presented in Figure~1 in \cite{L14} (for the case of an equal-mass
binary) and in Figure~5 in \cite{R14} (for the case of a
zero-eccentricity binary). Transitions (possibly involving
thresholds in the values of the problem parameters) from these two
limiting ``circular'' cases to the ``spiral'' one deserve a
separate study.

Another important issue that have not been addressed here, but
deserves a study, is the propagation of the secular spiral wave in
the presence of planets (which themselves can generate patterns in
the disk). Indeed, modern theories favor scenarios in which
planets form in gas-rich environments (see, e.g, the review by
\citealt{Z12}), i.e., the planets should be already present when
the secular spiral wave start to propagate.

Finally, note that the derived one-armed spiral pattern is a {\it
secular} one; i.e., it does not concern any orbital resonances. On
the other hand, the presence of multi-armed spiral patterns in
astrophysical disks is usually associated to Lindblad resonances;
see \cite{BT08} for the case of galactic dynamics and \cite{MD99}
for the case of planetary ring dynamics. Whether multi-armed
structures are possible to observe in the case of planetesimal
circumbinary disks? We think that this is not possible, because
all low-order resonances are deep inside the central chaotic zone
around the binary (if $\mu \gtrsim 0.05$, see \citealt{S15}).
Conversely, observable multi-armed structures can be generated by
resonances in the circumstellar case, i.e., in the inner disks,
because the stellar companions are inside their individual
stability regions.

\bigskip


We are grateful to the referee for useful remarks. We wish to
thank Alice Quillen for useful remarks and comments, and Natalia
Sotnikova for providing the SPH-code software. This work was
supported in part by the Russian Foundation for Basic Research
(projects Nos.\ 14-02-00319 and 14-02-00464) and the Programme of
Fundamental Research of the Russian Academy of Sciences
``Fundamental Problems of the Solar System Study and
Exploration''.

\newpage

\end{document}